
\magnification 1200
\baselineskip 22pt
\def\G{$\gamma$}
\def\g{\gamma}
\def\e{\eta}
\def\ee{$\eta$ }
\def\sles{\lower2pt\hbox{$\buildrel {\scriptstyle <}
   \over {\scriptstyle\sim}$}}
\def\sgreat{\lower2pt\hbox{$\buildrel {\scriptstyle >}
   \over {\scriptstyle\sim}$}}
\bigskip
\bigskip
\centerline{\bf STABILITY OF FIREBALLS AND $\gamma$-RAY BURSTS.}
\bigskip
\centerline{Eli Waxman}
\centerline{Institute for Advanced Study, Princeton NJ 08540, USA}
\medskip
\centerline{\it and}
\medskip
\centerline{Tsvi Piran}
\centerline{Racah Institute for Physics, The Hebrew University}
\centerline{Jerusalem, 91904, Israel}
\bigskip
\bigskip
\centerline{\bf Abstract}
Fireballs are an essential part of any cosmological $\gamma$-ray
burst. We derive here a stability criterion for fireballs and show that
fireballs are Rayleigh-Taylor unstable in any region in which the
entropy decreases outward. The instability begins to operate when the
fireball becomes matter dominated. Among the possible implication of
the instability are: (i) Conversion of a fraction of the radiation
energy to a convective energy expressed in the motion of bubbles
relative to each other.  (ii) Penetration of fast bubbles through
slower ones and creation of high $\gamma$ regimes which are essential
for efficient conversion of the energy to $\gamma$-rays.  (iii)
Formation of rapid variation (of the scale of the bubbles) in the
observed $\gamma$-rays.
\vfill\eject
\centerline{\bf 1. Introduction: Fireballs and Grbs}
\nobreak

Regardless of the nature of their source, cosmological $\gamma$-ray bursts
(GRBs) inevitably involve the creation of an expanding ``fireball'':
an extremely optically thick plasma of electrons and  positrons.
Even if the energy is initially entirely in the form of
gamma-radiation, electrons and positrons will be spontaneously created
by the reaction $\gamma+\gamma\rightarrow e^++e^-$ (Schmidt 1978,
Cavallo \& Rees 1978). The understanding of fireball evolution is therefore
crucial to the understanding of the observed GRBs.

The current picture of fireball evolution is based on smooth spherical
models (Goodman, 1986, Paczy\'nski 1986, Shemi \& Piran, 1990, Piran,
Narayan \& Shemi, 1993, M\'es\'zaros and Rees 1992, Rees \&
M\'es\'zaros, 1992) with a possible inclusion of shocks (Narayan,
Paczy\'nski \& Piran, 1992; Rees \& M\'es\'zaros 1994). In this letter
we show that fireballs may be Rayleigh-Taylor unstable.  The
instability develops during and after the transition from radiation
dominated to matter dominated fireball and therefore causes the
evolution to differ from that predicted by a smooth spherical
model. The existence of the instability may ease the constrains on the
baryonic rest mass allowed to be entrained within the fireball, and
may also have direct observational implications.

\bigskip
\centerline{\bf 2. Fireball Evolution, Asymptotic Laws and Shocks Waves}
\nobreak

Piran, Shemi \& Narayan (1993) have analyzed the evolution of a
fireball in terms of the evolution of its individual shells. This
descriptions holds for an impulsive fireball, in which all the energy
is released instantaneously (Goodman, 1986), as well as for a
quasi-steady radiative wind, in which the energy is released over a
period longer than the dynamical time (Paczy\'nski, 1986). Under certain
conditions, it is also valid for non spherical fragments of a
spherical shell (Piran, 1994; Narayan \& Piran, 1994).

Each shell is characterized by a radiation energy density $e_r$, a baryon
number density $n$ and a relativistic Lorentz factor \G. The evolution
may be divided into two phases. Initially the shell is radiation dominated,
$\e \equiv e_r/m_pnc^2 \gg 1 $, and the evolution is described by: $e_r
\propto R^{-4}$, $n \propto R^{-3}$, $\g \propto R$ and $\e \propto R^{-1}$
($R$ is the fireball radius). This description is valid after a rapid
initial acceleration that gives the shell relativistic velocity. Since
$\eta$ decreases the shell eventually becomes matter dominated. In the
limit of $\e \ll 1$ the baryons coast with: $\g \approx \e_i$ (the
index $i$ denotes initial quantities) and $n \propto R^{-2}$. The
radiation continues to cool with $\e \propto R^{-2/3}$. The transition
between radiation and matter dominated evolution occurs roughly at
$R=R_\eta\equiv R_i\eta_i$, where $\e\approx 1$. $\eta$ and $\gamma$
are related by $\gamma(4\eta/3+1)-
\eta/3\gamma=\eta_i+1$ throughout the shell evolution.

The fireball becomes optically thin (at $R=R_\tau$) during the
radiation dominated phase only if $\e$ is extremely large (typically
$>10^5$). In this case the released radiation is thermal with $T
\approx T_i(R/R_i)$ (Due to the relativistic expansion an observer at
rest will see blue shifted radiation with $T_{obs} \approx \g T
\approx T_i$).  If $\e_i$ is substantially smaller the shell becomes
matter dominated ($\e\approx 1$) before it becomes optically thin, and
only an insignificant amount of radiation energy is left at $R\approx
R_\tau$. However, not everything is lost. At a much larger radius the
baryons may interact with the ISM to produce a SNR like shock
(M\'es\'zaros \& Rees 1993; M\'es\'zaros, Laguna, \& Rees, 1993) The
kinetic energy is then converted back to radiation at the shock and it
produces a GRB if $\g \sgreat 10^3$. The last condition poses a strong
limit on the baryonic load: $M\ \sles\ 10^{-6} M_\odot (E/10^{51}{\rm
ergs})$, which might be difficult to satisfy.  This stringent limit is
the most serious open question that confronts cosmological GRB models
today.

This simple description breaks down when the model predicts shell
crossing.  Consider two adjoint shells denoted $h$ and $l$ with $R_h <
R_l$ separated initially by a distance $\Delta r$.  When the shells
become matter dominated they coast at $\g_h \approx \e_{i_h}$ and
$\g_l \approx \e_{i_l}$.  If $\e_{i_l} < \e_{i_h}$ the inner shell
moves faster and overtakes the outer one at $R_{\e_l}+
\Delta R_{sc} \approx R_{\e_l} + \eta_l^2 \Delta r$.  $\Delta R_{sc} $
is almost independent of $\eta_h$ (provided that $\eta_h\gg\eta_l$).
Shell crossing signifies the breakdown of the simple description given
in the previous paragraphs (in fact it breaks down slightly earlier)
and the formation of shock waves.  In general, the expansion is smooth
if $d\e_i /dr >0$ and shocks appear if $d\e_i /dr <0$.  Narayan
et. al.  (1992) and recently Rees \& M\'es\'zaros (1994) suggested
that such shocks could convert a fraction of the kinetic energy back
to radiation and, if this takes place at $R>R_\tau$, produce GRBs even
with a modest \ee fireballs.

\bigskip
\centerline{\bf 3. Stability Analysis}
\nobreak

\centerline{\bf 3.1 Short Wavelength Limit and the Isobaric Approximation}
\nobreak

Large scale deviations from spherical symmetry lead, as mentioned in
\S 2, to quasi-spherical local expansion.  We shall therefore be interested
only in the short wave-length limit in which the transversial linear scale of
the perturbations, $L$, is small compared with the fireball's thickness
$R/\g$.

The baryon's thermal energy is always negligible in a fireball. Hence
the pressure is radiation dominated, $p=aT^4 /3$, and the proper
energy density is: $e=nmc^2+aT^4$. The speed of sound,
$c^2_s=c^2/3(1+3/4\eta)$, is therefore comparable to the speed of
light as long as $\eta$ is not much smaller than unity.  The time
required to restore pressure equilibrium in a perturbed fluid element
of size $L$ (in the fluid's rest frame) is $\sim L/c_s\sim L/c$. In
the observer frame it equals $\gamma L/c$. For $L<R/\gamma$ this is
shorter than the time for changes in the unperturbed solution, $\sim
R/c$.  Thus, for $L<R/\gamma$ the perturbations are isobaric, i.e. we
may assume that the pressure perturbation vanishes.  (This assumption
holds provided that the typical growth rate of the instability is not
much faster than the typical rate of changes in the unperturbed
solution).

\bigskip
\centerline{\bf 3.2 Displacement Evolution }
\nobreak

We analyze the time evolution of a small fluid element displaced from
its original trajectory, $r_0(t)$.  Quantities along $r_0(t)$ are
denoted by $0$ subscripts ($\gamma_0(t)$, $p_0(t)$, $e_0(t)$ etc...), while
quantities along the perturbed trajectory $r'(t)$ are primed. The
displacement of the element is defined as $\delta r\equiv r'-r_0$.
Keeping only linear terms in $\delta r$ we obtain the acceleration
of the displacement, $\delta\ddot r$:
$$
\eqalignno{
\delta\ddot r={c^2\over \gamma^2_0\left(e_0+p_0\right)} \Biggl\{
&\ \left[\left(1+\left({\partial e\over\partial p}\right)_S\right)
{1\over e_0+p_0}\left({\partial p\over\partial r}\right)^2-
{\partial^2p\over\partial r^2}\right]\delta r\cr
&+{1\over c^2}\dot r_0\left[\left(1+\left({\partial e\over\partial p}\right)_S
\right){1\over e_0+p_0}{\partial p\over\partial t}{\partial p\over\partial r}
-{\partial^2p\over\partial r\partial t}\right]\delta r\cr
&+2{1\over c^2}\dot r_0\gamma_0^2\left({\partial p\over\partial r}+
{1\over c^2}\dot r_0{\partial p \over\partial t}\right)\delta\dot r
-{1\over c^2}{\partial p\over\partial t}\delta\dot r \Biggr\}.&(4)
}
$$
Here, all pressure derivatives are evaluated along the unperturbed
trajectory, $(r=r_0(t),t)$, and $S$ is the entropy.

In the limit of a non relativistic flow with a non relativistic
equation of state Eq. (4) reduces to the equation derived using a
somewhat different method by Goodman (1990).  For an incompressible
fluid, $c_s\rightarrow\infty$, in a steady state where pressure
gradient supports the fluid against gravity, $\rho^{-1}\partial
p/\partial r=-g$, Eq. (4) gives the classical Rayleigh-Taylor
stability criterion, $\partial\rho/\partial r<0$.

\bigskip
\centerline{\bf 3.3 A Qualitative Criterion for Local Convective Stability}
\nobreak

Eq. (4) describes the behavior of the Lagrangian displacement of a
perturbed fluid element. The growth of the displacement does not
indicate, however, that a convective instability
exists. Consider, for example, the flow of an expanding sphere where
the velocity increases outward. If a fluid element is displaced from
its original shell to some outer shell and continues to move with the
outer shell, the displacement will grow although the fluid element
does not overtake outer shells, i.e. although the instability is not
convective.

Convective instabilities exist only if the acceleration experienced by
the displaced fluid element is larger than the unperturbed
acceleration of the fluid at the new position of the fluid element
(rather than that at the original position of the element).  The
difference between the acceleration of a displaced fluid element and
the unperturbed acceleration of the fluid in the displaced position at
the rest frame of the shell into which the element had been displaced
is
%
%
$$
\delta a={c^2\over e+p}{\partial p\over\partial r}\ {1\over e+p}
         \left[\left({\partial e\over\partial p}\right)_S
         {\partial p\over\partial r}- {\partial
         e\over\partial r}\right]\delta r= {c^2\over
         (e+p)^2}\left({\partial p\over\partial r}\right)^2
         \left[\left({\partial e\over\partial p}\right)_S- {{\rm
         d} e\over{\rm d}p}\right]\delta r, \eqno(5)
$$
where the derivatives are evaluated at the shell rest frame and ${{\rm
d} e/ {\rm d} p}\equiv[(\partial e/\partial r )/(\partial p /\partial
r)].  $ Using $e=nmc^2+aT^4$ and $p=aT^4/3$, eq. (5) may be written in
the form
$$
\delta a={c^2\over (e+p)^2}{3\over\eta}p \left[{\partial
S\over\partial r}{\partial p\over\partial r}\right]_{RF}
\delta r, \eqno(6)
$$
where S, the entropy per baryon, is ${\rm log}\,T^3/n$. The subscript $RF$
denotes that the derivatives are to be taken at the shell rest frame.

Eq. (6) implies that convective instability will occur whenever the
rest frame pressure and entropy gradients are parallel. This criterion
is similar to the Schwarzschild criterion for convection is
stars. However, at the case at hand, the fluid background flow is not
stationary. Thus, in order for the instability to be of importance it
is necessary that the time scale for instability growth be comparable to
or shorter than the flow dynamical time scale. Only than a true dynamical
instability will exist. We therefore compare the typical time
in the observer frame for instability growth, $\tau_i\sim\gamma\sqrt
{\delta r/\delta a}$, with the typical time for shell acceleration,
the dynamical time $\tau_\gamma$,
$$\tau_{\gamma}\equiv{\gamma\over\dot\gamma}= {\gamma\over\gamma^3
v\dot v/c^2}=
\gamma {(e+p)\over c}\left({\partial p\over\partial r}\right)^{-1}_{RF}.$$
The ratio of these times is
$$\Theta\equiv{\tau_i\over\tau_\gamma}\sim
\sqrt{\eta\over 3}\left[{\partial\log p/\partial r\over\partial S/
\partial r}\right]^{1/2}_{RF}.\eqno(7)$$

We conclude that: (i) A fireball flow is convectively unstable in
regions where the entropy per baryon decreases outward (the pressure
is assumed to decrease outward).  (ii) As long as the fireball is
highly dominated by radiation, $\eta\gg 1$, the typical time for
convective instability development is large compared to the dynamical
time, $\Theta\gg 1$. However, as the shell approaches the matter
dominated phase, i.e. when $\eta\sim\rm few$, convective instabilities
will develop rapidly. A rapid instability development may be obtained
for larger $\e$ values if the entropy scale length is considerably
smaller than the pressure scale length.

The displaced fluid elements are accelerated by radiation pressure.
The energy driving the instability is therefore the radiation energy.
At the stage when the instability is expected to develop rapidly,
i.e.  when $\eta\sim\rm few$, most of the shell energy is in the
radiation field and had not been transformed to baryon kinetic
energy. This energy is therefore available for driving the
instability. In the limit of $\eta\ll 1$ most of the radiation energy
had already been converted to baryon energy, and only a small fraction
of it is available for the instability. At this stage the instability
will continue to develop due to the energy it acquired during the
transition phase from radiation to matter dominated evolution.
However, the energy associated with the instability will not increase
substantially during this phase (The displacement will continue to
grow, however without substantial acceleration).

\bigskip
\centerline{\bf 4. The Competition between Instability Growth and Shock
Formation}
\nobreak

Shock waves are formed in the (Lagrangian) fireball regions where $\partial
\eta_i/\partial r<0$. Unless the energy density increases steeply with $\e$,
such regions will also have $\partial S/\partial r<0$. This implies
that regions where shocks will develop, i.e. where
$\partial\eta_i/\partial r<0$, are quite generally also convectively
unstable. As explained in \S 2, a shock wave is formed at
$R_{sc}=R_\eta+\eta^2\Delta r$ as the inner high $\eta$ shells
overtake the outer low $\eta$ shells ($\Delta r$ is the initial
separation of the shells). The instability, on the other hand, begins
to operate at $R\sim R_\e$. Thus, if $\Delta r\gg R_i /\eta$ then
$R_{sc}-R_\e\gg R_\e$ and the shock forms long after the instability
started developing. If, however, the inverse condition $\Delta r \sles
R_i /\eta$ holds, instability development and shock formation are
competing processes.  The following numerical example demonstrates
that even in this case the instability develops substantially prior to
shock formation.

We have calculated numerically the evolution of a fireball with an
initial radius $R_i=10^7 \rm cm$, initial temperature $T_i=1 \rm MeV$,
and $\eta_i$ profile decreasing outward from $\eta_h=50$ to
$\eta_l=5$ over $\Delta r\sim 7\cdot 10^5\rm cm$. Fig. 1 presents the
solution of (4) for the evolution of the displacement of a fluid
element initially lying at the $\eta_i$ transition layer (we have
chosen $\delta r>0$ and $\delta\dot r=0$ at $t=t_i$). The displacement
grows by the time of shock formation, $r_0/R_i\approx 30$, by two
orders of magnitude. The instability is indeed convective, as seen by
comparing the displacement evolution to the evolution of the distance
between two adjoint shells (Note that the initial rapid displacement
growth, obtained for $r_0/R_i\sles 1$, is not convective and is due to
the differential expansion). The results presented in fig. 1 are
consistent with the qualitative analysis of \S 3.3, according to which
fireball regions where $\partial S/\partial r<0$ are subject to
substantial convective instability growth during the transition from
radiation to matter dominated evolution.
A shock wave forms only at  the end of the
computation shown in fig. 1. This demonstrates that the instability
indeed won over the shock formation.  It should be noted that as the
inner shell overtakes the outer shell some of the baryon kinetic
energy is converted back to radiation energy as the pressure in the
shells interface region increases. This energy is then available for
driving the instability.

\bigskip
\centerline{\bf 5. Conclusions and Applications to Astrophysical Fireballs
and Grbs}
\nobreak

We have seen that a convective instability develops whenever $\partial S/
\partial r< 0$. This includes, quite generally, the important case of
$\partial\e /\partial r < 0$, which appears whenever an inner high
$\eta$ shell accelerates an external low $\e$ shell. The instability operates
from the stage of transition from radiation to matter dominated evolution,
$\eta\sim\rm few$ and $R\sim R_\e$, until the fireball
becomes optically thin at $R_\tau$ (or at $\eta^{1/2} R_\tau$ where
the electrons decouple from the photons). The instability competes
with shocks that form at shell crossing within $ \Delta R_{sc} $ after
$R_\eta$.

We turn now to an astrophysical fireball which appears in a
cosmological $\gamma$-ray burst.  The energy release is $10^{51}$ergs
and the rising time indicates that the initial radius $R_i$ is
around $10^7$cm (or less). With these parameters we have: $R_\e=
R_i \e_i = 10^{11}{\rm cm} R_{i7} \e_{i4}$ (where $R_{i7} =R_i/10^7$cm and
$\e_{i4}=\e_i/10^4$), $R_\tau= (\sigma_T E /\e_i m_p c^2 )^{1/2} = 6\cdot
10^{12}{\rm cm} E^{1/2}_{51} \e^{-1/2}_{i4}$ (where
$E_{51}=E/10^{51}$ergs) and $\Delta R_{sc} \approx \Delta r \eta_l^2
=10^{14}{\rm cm} \Delta r_6 \e_{l4}^2 = 10^3 R_\e (\Delta r_6 /R_{i7})
\e_{l4}$. Thus, for $\e< 3.3 \cdot 10^4 E_{51}^{1/3} R_{i7}^{-2/3}$ the
instability will operate during a period when the fireball expands by
a factor of ten or more.  This gives ample time for the growth of the
instability.  Shell crossing and shocks occur within $\Delta R_{sc}$ from
$R_\e$. If $\Delta r$ is not too small then $\Delta R_{sc} \gg R_\e$.
However, even when this condition is not satisfied we have demonstrated that
the instability operates rapidly enough before the shock forms. We
conclude that generally, the instability has enough time to develop
substantially whenever the conditions for its existence are
fulfilled.

We stress that our  stability analysis presented  is
valid for a quasi-steady wind flow generated when the energy is
released over a time scale larger than the light crossing time of the
source, as well as for a fireball flow where the energy release is
instantaneous. The (Lagrangian) regions of a wind created during
periods when the emmited entropy increases with time, leading to a
spatial profile with $\partial S/\partial r<0$, will be unstable.
This might have important implications to the recent model, suggested
by Rees \& M\'es\'zaros (1994), according to which GRBs are produced
due to the formation of shocks between adjoint shells in the
fireball. The formation of such shocks requires a wind where $\eta$
increases with time, resulting in regions with $\partial\eta/\partial
r<0$. Such regions are likely to satisfy $\partial S/\partial r<0$
(unless the source luminosity varies enormously with $\e$) and are,
therefore, unstable.  This implies that the effects of the instability
must be taken into account within the framework of this model.

The convective fireball instability described here will give rise to
convection zones where high $S$ bubbles penetrate into low $S$ shells
and vice versa. It is difficult to estimate, at present, the full
implications of the instability to fireballs and to the observation of
$\gamma$-ray bursts, since for such an estimate it is necessary to
consider the non-linear regime of instability development, where
non-linear dissipation affects bubble motion.  In the following we
list several possible effects.  First and most important is the fact
that a situation where a high $\eta$ shell accelerates a low $\eta$
shell is unstable.  The instability may therefore provide a mechanism
for allowing high $\eta$ shells to penetrate through low $\eta$ ones,
thus allowing to achieve Lorentz factors higher than that
corresponding to the average $\eta$. This may ease the stringent limit
on the allowed baryonic load.

The mixing induced by the instability of high and low $\eta_i$ shells
will give rise to regions where the baryon density $n$ fluctuates over
length scales much shorter than both the fireball thickness and the
source size, which are comparable and of the order of $R_i$. If the
burst is produced by a collision with the ISM or by collisions of
sub-shells within the fireball, this small scale structure may account
for the rapid temporal variations of the observed pulse. This effect
is especially important for the scenario where the burst is produced
by a collision with the ISM. In this case, the duration over which an
observer receives the radiation from a single sub-shell of the
fireball is (Katz 1994) $R_c/\eta_i c$, where the collision radius
$R_c$ is (Piran 1994) $R_c=1.3\cdot 10^{15}{\rm
cm}E^{1/3}_{51}\eta^{-2/3}_{i4} n^{-1/3}_I$ ($n_I$ is the ISM number
density in units of ${\rm cm}^{-3}$).  This duration is much larger
than the source dynamical time, $R_c/\eta_i R_i= 1.3\cdot 10^4
E^{1/3}_{51}\eta^{-5/3}_{i4}n^{-1/3}_I$. Thus, it would be hard to
explain the rapid temporal variations of the pulse as resulting from
temporal variations of the source.

Finally, the generation of local random motions within the fireball shell
will cause the fireball evolution to differ from that of a spherical
smooth model due to the mixing induced by these motions and due to the
kinetic energy associated with them. Such motions may further contribute to
the build up of a magnetic field, which is necessary for reconverting the
baryon kinetic energy to radiation energy.

We thank Ramesh Narayan for helpful discussions.
This research was partially supported by a BRF grant to the Hebrew
University, and by NSF grant PHY 92-45317 and W. M. Keck Foundation
grant to the Institute for Advanced Study.

\bigskip
\centerline{\bf References}
\nobreak

\def\ApJ{{\it Ap. J.}}
\def\ApJL{{\it Ap. J. Lett.}}

\item{}
Cavallo, G. and Rees, M.J., 1978, {\it MNRAS.}, {\bf 183}, 359
\item{}
Goodman, J., 1986, \ApJL, {\bf 308} L47.
\item{}
Goodman, J., 1990, \ApJ, {\bf 358}, 214.
\item{}
Katz, J. I., 1994, \ApJ, {\bf 422}, 248.
\item{}
M\'es\'zaros, P., \& Rees, M. J.,  1992.  {\it MNRAS.}, {\bf 257}, 29p.
\item{}
M\'es\'zaros, P., \& Rees, M. J.,  1993.  \ApJ, {\bf 405}, 278.
\item{}
M\'es\'zaros, P., Laguna, P., \& Rees, M. J.,  1993.  \ApJ, {\bf 415}, 181.
\item{}
Narayan., R., Paczy\`nski, B., \& Piran, T., 1992, \ApJL, {\bf 395}, L83.
\item{}
Rees, M. J., \& M\'es\'zaros, P., 1992.  {\it MNRAS.}, {\bf 258}, 41p.
\item{}
Rees, M. J., \& M\'es\'zaros, P., 1994.  astro-ph/9404038.
\item{}
Paczy\'nski, B., 1990. \ApJ, {\bf 363}, 218.
\item{}
Piran, T., Shemi, A. and Narayan, R., 1993, {\it MNRAS.}, {\bf 263}, 861.
\item{}Piran, T., 1994, to appear in the Proceedings of the second
Huntsville GRB conference, AIP press.
\item{}
Schmidt, W. K. H., 1978, Nature, {\bf 271}, 525.
\item{}
Shemi, A. and Piran, T. 1990, \ApJL {\bf 365}, L55.

\bigskip
\centerline{\bf Figure Captions}
\nobreak
Fig. 1: The evolution of a fireball sub-shell initially lying at the
$\e_i$ transition layer.  The horizontal axis measures the distance
which the shell moved, which is proportional to the time that passes
from the beginning.  Solid line: The displacement $\delta r$
normalized to its initial value $\delta r_i$; Dashed line: $\e$;
Dash-dot line: $\Theta$; Dotted line: The distance between the shell
and an outer adjoint shell normalized to the distance at $t_i$.
Note that a shock wave  forms only at  the end of this
computation.

\end